\documentclass[usegraphicx,usenatbib,useAMS,letterpaper]{mn2e}

\usepackage{amssymb}
\usepackage{rotating}

\newcommand{\apj}{ApJ}

\newcommand{\apjs}{ApJS}
\newcommand{\aj}{AJ}
\newcommand{\aap}{A\&A}

\newcommand{\aaps}{A\&AS}
\newcommand{\araa}{ARA\&A}
\newcommand{\pasj}{PASJ}

\newcommand{\mnras}{MNRAS}
\newcommand{\pasp}{PASP}
\newcommand{\apss}{Ap\&SS}

\newcommand{\jpb}{J. Phys. B: At. Mol. Opt. Phys.}

\newcommand{\eas}{EAS Pub. Series}

\newcommand{\subscript}[1]{\textnormal{\scriptsize{#1}}}
\newcommand{\rdout}{\ensuremath{R_\subscript{out}}}
\newcommand{\rdin}{\ensuremath{R_\subscript{in}}}

\newcommand{\rstar}{\ensuremath{R_\star}}

\newcommand{\kms}{km~s$^{-1}$}
\newcommand{\cm}{cm$^{-1}$}

\newcommand{\irc}{IRC+10216}

\begin{document}

\title[SiS in the innermost envelope of \irc]
{The abundance of $^{28}$Si$^{32}$S, $^{29}$Si$^{32}$S, $^{28}$Si$^{34}$S, and 
$^{30}$Si$^{32}$S in the inner layers of the envelope of \irc}
\author[J. P. Fonfr\'{\i}a et al.]{
J. P. Fonfr\'ia,$^1$\thanks{fonfria@astro.unam.mx} 
J. Cernicharo,$^2$ 
M. J. Richter,$^3$\thanks{Visiting Astronomer at the Infrared Telescope 
Facility, which is operated by the University of Hawaii under contract from 
the National Aeronautics and Space Administration.} 
M. Fern\'andez-L\'opez,$^4$
\newauthor 
L. Velilla Prieto$^2$
and J. H. Lacy$^5$\footnotemark[2]\\
$^1$Departamento de Estrellas y Medio Interestelar, Instituto de Astronom\'ia, 
UNAM, Ciudad Universitaria, 04510, Mexico City (Mexico)\\
$^2$Grupo de Astrof\'isica Molecular, Instituto de Ciencia de Materiales
de Madrid, CSIC, C/ Sor Juana In\'es de la Cruz 3, 28049,\\
Cantoblanco, Madrid (Spain)\\
$^3$Physics Dept. - UC Davis, One Shields Ave., Davis, CA 95616 (USA)\\
$^4$Instituto Argentino de Radioastronom\'ia, CCT-La Plata (CONICET), C.C.5,
1894, Villa Elisa (Argentina)\\
$^5$Astronomy Dept., University of Texas, Austin, TX 78712 (USA)}

\maketitle

\begin{abstract}
We present high spectral resolution mid-IR observations of SiS towards the 
C-rich AGB star \irc{} carried out with the Texas Echelon-cross-Echelle 
Spectrograph mounted on the NASA Infrared Telescope Facility.
We have identified 204 ro-vibrational lines of $^{28}$Si$^{32}$S,
26 of $^{29}$Si$^{32}$S, 
20 of $^{28}$Si$^{34}$S, and
15 of $^{30}$Si$^{32}$S
in the frequency range $720-790$~\cm.
These lines belong to bands $v=1-0$, $2-1$, $3-2$, $4-3$, and $5-4$,
and involve rotational levels with $J_\subscript{low}\lesssim 90$.
About 30~per~cent of these lines are unblended or weakly blended and can be 
partially or entirely fitted with a code developed to model the mid-IR 
emission of a spherically symmetric circumstellar envelope composed of 
expanding gas and dust.
The observed lines trace the envelope at distances to the star 
$\lesssim 35$\rstar{} ($\simeq 0\farcs7$).
The fits are compatible with an expansion velocity of 
$1+2.5(r/\rstar-1)$~\kms{} between 1 and 5\rstar, 11~\kms{} between 5 and
20\rstar, and 14.5~\kms{} outwards.
The derived abundance profile of $^{28}$Si$^{32}$S with respect to H$_2$ is
$4.9\times 10^{-6}$ between the stellar photosphere and 5\rstar, 
decreasing linearly down to $1.6\times 10^{-6}$ at 20\rstar{} and to
$1.3\times 10^{-6}$ at 50\rstar.
$^{28}$Si$^{32}$S seems to be rotationally under LTE in the region of the 
envelope probed with our observations and vibrationally out of LTE in most of 
it.
There is a red-shifted emission excess in the $^{28}$Si$^{32}$S lines of 
band $v=1-0$ that cannot be found in the lines 
of bands $v=2-1$, $3-2$, $4-3$, and $5-4$.
This excess could be explained by an enhancement of the vibrational
temperature around 20\rstar{} behind the star.
The derived isotopic ratios $^{28}$Si/$^{29}$Si, and 
$^{32}$S/$^{34}$S are 17 and 14, compatible with previous estimates.
\end{abstract}

\section{Introduction}
\label{sec:introduction}

\irc{} (CW Leo) is the closest asymptotic giant branch (AGB) star to Earth
\citep*[$\simeq 120$~pc;][]{groenewegen_2012}.
During the last decades this star was usually considered as an isolated star 
with an effective temperature of $\simeq 2300$~K \citep{ridgway_1988} and
surrounded by an optically thick C-rich circumstellar envelope (CSE) composed 
of molecular gas and dust, which results from the ejection of stellar material 
at a rate $\simeq 1-5\times 10^{-5}$ 
\citep*[e.g.,][]{knapp_1985,cernicharo_1996,schoier_2001,debeck_2012}.
Although there were indirect evidences suggesting that it could be actually a 
binary system 
\citep*[][]{mauron_1999,mauron_2000,fong_2003,decin_2011,decin_2014,
cernicharo_2015a,cernicharo_2015c},
\citet{kim_2014} recently reported the discovery of what could be a faint 
companion of the AGB star responsible of the ejected matter.

More than 80 molecular species have been detected in its CSE
\citep*[e.g.,][]{kawaguchi_1995,cernicharo_2000,he_2008,patel_2011}.
Although most of them form in the outer envelope, the so-called parent 
molecules (e.g., CO, C$_2$H$_2$, HCN, SiS, SiO, CS, SiC$_2$, Si$_2$C) 
are formed in warm and dense shells near the stellar photosphere
\citep{keady_1993,boyle_1994,cernicharo_2010,cernicharo_2015b}, 
where strong dynamical 
mechanisms probably exist \citep{hinkle_1982,bowen_1988} and an active 
chemistry has been inferred \citep{willacy_1998,agundez_2006,cherchneff_2006}.
The dust formation and the subsequent gas acceleration involved in the 
development of the envelope 
\citep*[e.g.,][]{gilman_1972,morris_1987,bowen_1988,fonfria_2008} 
are supposed to be
intimately related to the refractory nature and abundance of the parent molecules, 
in particular to C- and Si-bearing species such as SiS, SiC$_2$,
and Si$_2$C \citep{cernicharo_2015b}.
These molecules are believed to be highly depleted due to dust throughout
the first $20-30$\rstar{} outwards from the stellar photosphere
\citep{turner_1987,bieging_1989,bieging_1993,boyle_1994,lucas_1995,agundez_2012,fonfria_2014}.

The emission of SiS has been analysed in the mid-infrared \citep{boyle_1994}, 
the submillimetre \citep{patel_2009,patel_2011,decin_2010,decin_2014}, 
and the millimetre frequency ranges 
\citep{morris_1975,henkel_1983,henkel_1985,turner_1987,
kahane_1988,kahane_2000,bieging_1989,carlstrom_1990,bieging_1993,lucas_1995,
lucas_1997,cernicharo_2000,cernicharo_2014,fonfria_2006,
schoier_2007,agundez_2012,fonfria_2014,velilla_2015}.
Its abundance throughout the middle and outer envelope is reasonably well 
estimated ($\simeq (1-2)\times 10^{-6}$ outwards from 
$\simeq 10-15\rstar\simeq 0.2-0.3$~arcsec) but its value in the innermost 
envelope is still controversial due to discrepancies of up to one order of 
magnitude between the results of several works.

The main goal of the current work is to derive a reliable SiS abundance profile
for the innermost envelope of \irc{} in order to settle the controversy
through the analysis of high spectral resolution molecular observations in the 
mid-IR, which has been proven as a powerful method to probe the region close 
to the central star \citep{keady_1988,keady_1993,boyle_1994,fonfria_2008}.

In this Paper we present the identification of 265 lines of the vibrational 
bands $v=1-0$, $2-1$, $3-2$, $4-3$, and $5-4$ of the silicon sulphide 
isotopologues $^{28}$Si$^{32}$S, $^{29}$Si$^{32}$S, $^{28}$Si$^{34}$S,
and $^{30}$Si$^{32}$S observed with high-spectral resolution ($\simeq 3-4$~\kms) 
towards \irc.
About 30~per~cent of them have enough quality to be properly fitted and 
analysed.
The observing procedure is included in Section~\ref{sec:observations}.
Comments on the fitting process of the molecular lines can be found
in Section~\ref{sec:modelling}.
The description of the observed SiS lines, the results of the fitting
procedure, and their analysis are addressed in Section~\ref{sec:results}.
Finally, a brief summary of the current work and our conclusions can be found 
in Section~\ref{sec:conclusions}.

\section{Observations}
\label{sec:observations}

We observed \irc{} with the Texas Echelon-cross-Echelle Spectrograph 
\citep*[TEXES;][]{lacy_2002} at the NASA Infrared Telescope Facility (IRTF) 
on 2002 Dec 12.
We used the TEXES high resolution echelon grating with a first order grating 
as the cross-disperser.
In this mode, a single setting covered a spectral range of $\simeq 0.25~\mu$m.
The entire range ($11.6-13.9~\mu$m) was covered with 10 separate settings.
We nodded \irc{} off the slit for sky subtraction. 
The weather was good enough during the observations that the median 
background level measured in the \textit{off} positions showed no significant 
variations.
We used a black body-sky difference spectrum to correct for the atmosphere.
The data were reduced with the standard TEXES pipeline \citep{lacy_2002}.
We normalised each spectrum by removing the baseline with a polynomial fit.
The rms of the random noise is $\simeq 0.15$~per~cent of the continuum emission. 

\begin{figure*}
\includegraphics[width=0.95\textwidth]{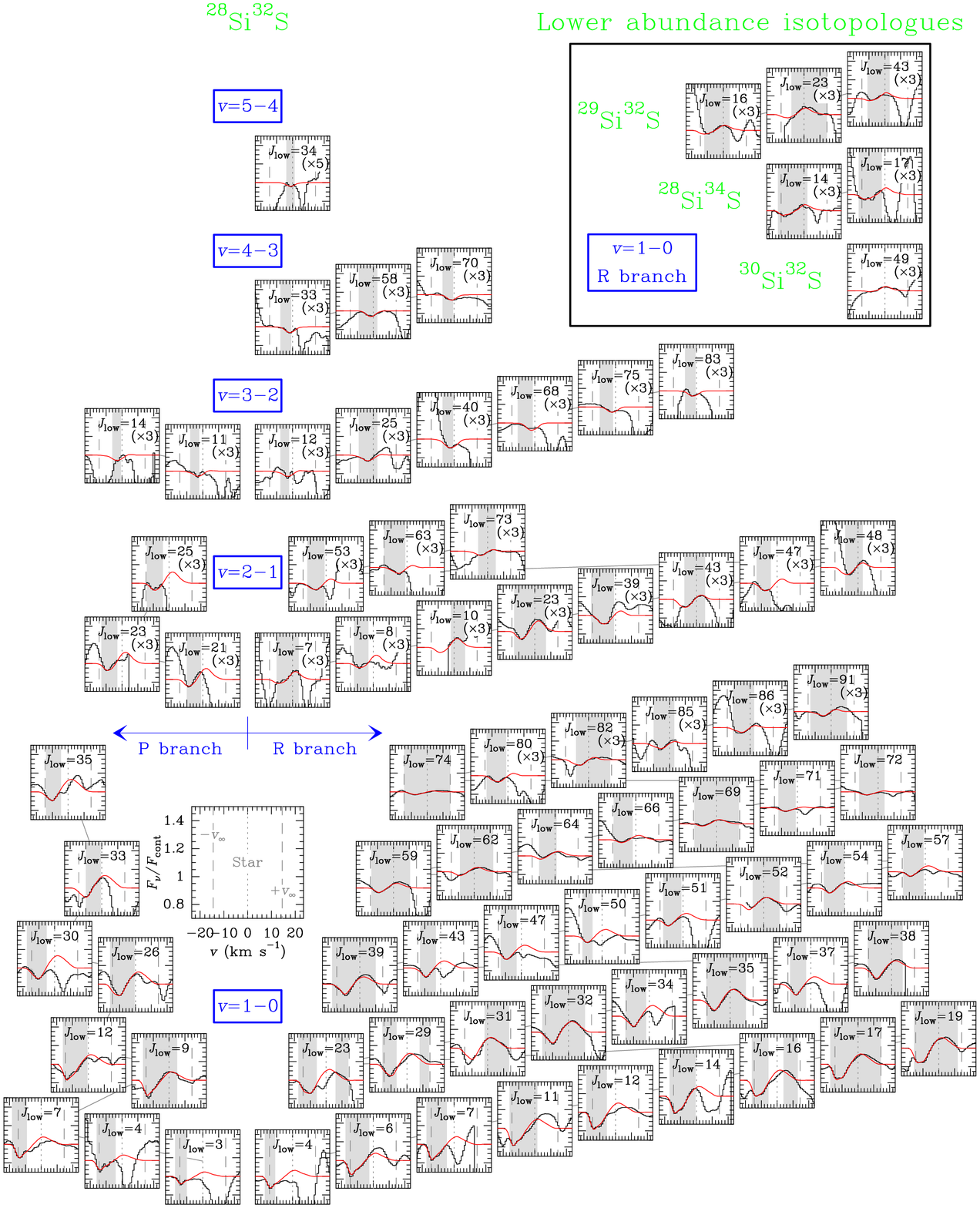}
\caption{$^{28}$Si$^{32}$S, $^{29}$Si$^{32}$S, $^{28}$Si$^{34}$S, and 
$^{30}$Si$^{32}$S lines in the observed spectrum (black histogram).
Most of the lines surrounding those of the SiS isotopologues belong to 
C$_2$H$_2$, HCN, and their isotopologues \citep{fonfria_2008}.
The synthetic spectrum is plotted in red.
The abscissa axis represents the Doppler velocity 
($v=\textnormal{V}_\subscript{LSR}-v_\subscript{sys}$, 
$v_\subscript{sys}=-26.5$~\kms; \citealt{cernicharo_2000,he_2008,patel_2011}).
The vertical dark grey dotted and dashed lines indicate the 
velocity of the star (0~\kms)
and the maximum and minimum gas expansion velocity 
projected on to the line-of-sight ($+v_\infty$ and $-v_\infty$), respectively
($v_\infty=14.5$~\kms).
The frequency range shaded in light grey in every inset indicates the
part of the lines used in the fits.
The intensity of some lines with respect to the baseline was multiplied
by 3 and 5 to improve their visibility
(it is indicated at the upper right corner of the corresponding
insets as $\times 3$ or $\times 5$).
}
\label{fig:f1}
\end{figure*}

Although most of the lines in the spectrum are produced by C$_2$H$_2$, HCN, 
and their main isotopologues \citep{fonfria_2008}, we have identified 204 
lines of $^{28}$Si$^{32}$S, 26 of $^{29}$Si$^{32}$S, 20 of $^{28}$Si$^{34}$S, and 
15 of $^{30}$Si$^{32}$S in the frequency range between 720 to 790~\cm.
A hundred and fifteen of the $^{28}$Si$^{32}$S lines belong to band $v=1-0$, 
54 to band $v=2-1$, 13 to band $v=3-2$, 16 of band $v=4-3$, and 6 of band 
$v=5-4$.
Seventy three of them are unblended or partially blended and can be fitted
(Fig.~\ref{fig:f1}).
The identified lines of $^{29}$Si$^{32}$S, $^{28}$Si$^{34}$S, and $^{30}$Si$^{32}$S 
pertain to band $v=1-0$.
Three, two, and one of them, respectively, are unblended or partially blended 
and can be reasonably fitted.
The lines of bands involving vibrational states with higher energy are below 
the detection limit.
No lines of the other SiS isotopologues, which are at least 100 times less 
abundant than $^{28}$Si$^{32}$S \citep*[e.g.,][]{kahane_2000,cernicharo_2000}, 
have been detected in our spectrum.

We have used the low spectral resolution \textit{ISO}/SWS observations 
carried out on 1996 May 31 \citep{cernicharo_1999} to determine the properties 
of the dusty envelope.
We estimate the uncertainty of these observations due to noise and the
calibration process in $\simeq 10$~per~cent.

\section{Model of envelope and fitting procedure}
\label{sec:modelling}

The observed SiS lines were fitted and analysed with the aid of the code 
developed by \citet{fonfria_2008}, which solves numerically the radiation 
transfer equation of a spherically symmetric envelope composed of gas and 
dust in radial expansion.
It was successfully used to analyse the mid-IR 
spectra of C$_2$H$_2$, HCN, C$_4$H$_2$, and C$_6$H$_2$ at 8 and $13~\mu$m 
towards \irc{} and the protoplanetary nebula CRL618 
\citep{fonfria_2008,fonfria_2011}.
The code \textit{does not} solve the statistical equilibrium equations,
which would require a very large amount of CPU time to determine the 
population of all the ro-vibrational levels involved in the calculations.
Instead, it makes use of rotational and vibrational temperature profiles 
provided by the user.
The code solves the radiation transfer equation and produces the spectrum 
of the envelope convolved with a given Gaussian beam.
The new 3D version of this code, presented in \citet{fonfria_2014},
allowed us to calculate as well the emission of a simple asymmetrical envelope.
A deeper insight on how this code works in its 1D and 3D versions
and its benchmarking can be found 
in \citet{fonfria_2008} and \citet{fonfria_2014}.

\begin{table}
\caption{Stellar and envelope parameters involved in the fits}
\label{tab:table1}
\begin{tabular}{c@{\hspace{0ex}}c@{\hspace{0ex}}c@{\hspace{0ex}}c@{\hspace{0ex}}c}
\hline
Parameter & Units & Value & Error & Ref.\\
\hline
$D$                               & pc                & 123                & ---           & 2 \\
$\alpha_\star$                     & arcsec            & 0.02               & ---           & 3 \\
\rstar                            & cm                & $3.7\times 10^{13}$ & ---           & \\
$\dot M$                          & M$_\odot$~yr$^{-1}$ & $2.1\times 10^{-5}$ & ---           & 4 \\
$T_\star$                          & K                 & $2330$             & ---           & 3 \\
$x_\subscript{He}$                  &                   & 0.2                & ---           & 5\\
$\rdin$                           & \rstar            & $5$                & ---           & 4 \\
$\rdout$                          & \rstar            & $20$               & ---           & 4 \\
$v_{\subscript{exp}}(1R_\star\le r<\rdin)^*$      & \kms              & $1+2.5(r/\rstar-1)$              & ---          &  1 \\
$v_\subscript{exp}(\rdin\le r<\rdout)$          & \kms              & $11.0$             & ---           & 4 \\
$v_\subscript{exp}(r\ge\rdout)$         & \kms              & $14.5$             & ---           & 4 \\
$T_\subscript{k}(1\rstar\le r< 9\rstar)$ & K             & $T_\star(\rstar/r)^{0.58}$  & ---     & 6   \\
$T_\subscript{k}(9\rstar\le r< 65\rstar)$ & K           & $T_\star(9\rstar/r)^{0.40}$ & ---     & 6   \\
$T_\subscript{k}(r\ge 65\rstar)$    & K                 & $T_\star(65\rstar/r)^{1.2}$ & ---     & 6   \\
$\Delta v(1\rstar\le r<\rdin)$          & \kms     & $5(\rstar/r)$       & ---           & 7 \\
$\Delta v(r\ge\rdin)$                  & \kms              & $1$                & ---           & 8 \\
$f_\subscript{AC}$                  & per~cent          & 95                 & ---           & 4 \\
$f_\subscript{SiC}$                 & per~cent          & 5                  & ---           & 4 \\
$\tau_\subscript{dust}(11~\mu$m)    &                   & 0.70               & $+0.11/-0.09$ & 1 \\
$T_\subscript{dust}(\rdin)$         & K                 & 825                & $+130/-160$   & 1 \\
$\gamma_\subscript{dust}$           &                   & 0.39               & $+0.04/-0.06$ & 1 \\
\hline
\end{tabular}
\newline
$D$: distance to the star; 
$\alpha_\star$: angular stellar radius;
\rstar: linear stellar radius;
$\dot M$: mass-loss rate; 
$T_\star$: stellar effective temperature;
$x_\subscript{He}$: abundance of He;
\rdin{} and \rdout: positions of the inner and outer acceleration shells;
$v_\subscript{exp}$: gas expansion velocity;
$T_\subscript{k}$: gas kinetic temperature;
$\Delta v$: line width;
$f_\subscript{X}$: fraction of dust grains composed of material X, 
i.e., amorphous carbon (AC) or silicon carbon (SiC);
$\tau_\subscript{dust}$: dust optical depth along the line-of-sight;
$T_\subscript{dust}$: temperature of dust grains;
$\gamma_\subscript{dust}$: exponent of the decreasing dust temperature power-law.
The parameters for which no uncertainty has been provided (---) were assumed
as fixed throughout the whole fitting process.
$^*$The fit to the observed SiS lines
was better with this expansion velocity field
in Region I and it was assumed during the whole fitting process
(see Section~\ref{sec:velocity}).
References: 
(1) This work
(2) \citet{groenewegen_2012}
(3) \citet{ridgway_1988}
(4) \citet{fonfria_2008}
(5) \citet{cox_2000}
(6) \citet{debeck_2012}
(7) \citet{agundez_2012}
(8) \citet{huggins_1986}.
\end{table}

\subsection{Adopted envelope model and spectroscopic constants}
\label{sec:model}

In order to analyse the SiS lines identified in the observed spectrum, we 
adopted the model of envelope composed of several concentric shells with a 
constant expansion velocity derived by \citet{fonfria_2008} from the analysis 
of the mid-IR spectra of C$_2$H$_2$ and HCN, which are produced in the same 
region of the envelope of \irc{} that the observed SiS lines trace.
Despite the fact that some authors used a different model with a more
continuous expansion velocity profile 
\citep*[e.g.,][]{schoier_2006b,decin_2010,decin_2014,debeck_2012},
the adopted model seems to be more suitable to describe the inner layers of 
the envelope 
\citep{keady_1988,keady_1993,boyle_1994,fonfria_2008,fonfria_2014,
cernicharo_2011,cernicharo_2013,agundez_2012,daniel_2012}.
With the aim of supporting our choice, we briefly discuss about the quality of 
the fits using the selected velocity profile and that recently derived by 
\citet{decin_2014} in Appendix~\ref{sec:expansion.velocity}.
Thus, we assumed that the CSE is divided into three Regions (I, II, and III) 
separated by two acceleration shells located at 
5 and 20\rstar, which 
were initially 0.5\rstar{} thick each (Table~\ref{tab:table1}).
The gas expansion velocity, $v_\subscript{exp}$, is initially assumed
to be 5, 11, and 14.5~\kms{} 
in Regions I, II, and III, respectively.
In order to improve the expansion velocity field close to the star,
different thicks of the inner acceleration shell (ending at 5\rstar)
and expansion velocities in Region I were tried.
The mass conservation results in a gas density profile following
very approximately the law $\propto r^{-2} v_\subscript{exp}^{-1}$, since
a negligible amount of the total ejected mass ($\lesssim 0.25$~per~cent)
is locked in the dust grains \citep{debeck_2012,decin_2014}.
We assumed that there is no atomic H in gas phase 
\citep{knapp_1983,glassgold_1996,agundez_2006}.
The line width at the stellar surface is assumed to be 5~\kms,
decreasing to the terminal value of 1~\kms{} at 5\rstar{}
following the power-law $\propto 1/r$
\citep*[e.g.,][]{huggins_1986,keady_1988,schoier_2006a,agundez_2012}.
This line width profile is compatible with the turbulent velocity 
$\gtrsim 3$~\kms{} suggested by \citet{keady_1988} and \citet{monnier_2000} 
and the line width of $\simeq 5-8$~\kms{} inferred from the vibrational excited 
lines of SiS, CS, and HCN observed in the mm and sub-mm wavelength ranges by 
\citet{patel_2009,patel_2011} and 
\citet{cernicharo_2011}, which are supposed to be formed 
in the inner layers of the envelope.
The kinetic temperature was adopted to follow the profile derived from the 
analysis of the CO emission by \citet{debeck_2012}, i.e., 
$T_\subscript{k}\propto r^{-0.58}$ between the stellar surface and 9\rstar, 
$\propto r^{-0.4}$ from 9\rstar{} to 65\rstar, and $\propto r^{-1.2}$ outwards.
The rotational and vibrational temperature profiles were assumed to follow 
$r^{-\gamma}$, a dependence on the distance to the star tightly related to the 
cooling process of the envelope \citep{mckee_1982,doty_1997}. 
$\gamma$ can be different for each region
and for the rotational and vibrational temperatures.  
$\gamma$ cannot be determined for both the rotational and vibrational 
temperatures in Region III due to the low sensitivity of the model to the 
emission coming from that zone.
Thus, we adopted $\gamma=0.55$ and 1.0,
respectively \citep{fonfria_2008,agundez_2012}.
We assumed SiS is in rotational LTE at the stellar photosphere.
No dust was considered to exist in Region I.
Throughout the rest of the envelope, we adopted dust grains composed of 
95~per~cent of amorphous carbon (AC) and 5~per~cent SiC, with a size of 
$0.1~\mu$m, and a density of 2.5~g~cm$^{-3}$ \citep{fonfria_2008}.

The rest frequencies of the observed lines 
and the energy of the vibrational states involved in the calculations 
($v\le 9$; $E_\subscript{vib}\lesssim 9400$~K)
were taken from the MADEX code \citep{cernicharo_2012} and the fits presented
by \citet{velilla_2015}.
The optical constants needed to calculate the opacity of the dust grains were
taken from \citet{rouleau_1991} for amorphous carbon and from 
\citet{mutschke_1999} for SiC.

\subsection{Fitting procedure}
\label{sec:fitting.procedure}

The fitting procedure is based on the minimisation of the reduced
$\chi^2$ function defined as
\begin{equation}
\chi_\subscript{red}^2=\frac{1}{n-p}\sum_{i=1}^n \frac{1}{\sigma_i^2}\left[
\left(\frac{F_\nu}{F_\subscript{cont}}\right)_\subscript{obs,$i$}-
\left(\frac{F_\nu}{F_\subscript{cont}}\right)_\subscript{synth,$i$}\right]^2,
\end{equation}
where $n$ and $p$ are
the number of frequency channels and of parameters
involved in the fitting process, 
$(F_\nu/F_\subscript{cont})_\subscript{obs}$ and
$(F_\nu/F_\subscript{cont})_\subscript{synth}$ are the flux normalised to the 
continuum for the synthetic and observed spectra, respectively,
and $\sigma_i$ is the uncertainty of the observed 
normalised flux for each channel.
In this work, $\sigma_i$ was considered equal to the noise rms for all the
frequency channels.
We assumed a number of physical parameters related to the envelope model
as fixed during the fitting process (gas expansion velocity field, H$_2$ 
density, kinetic temperature, or line width, among others; see 
Tables~\ref{tab:table1} and \ref{tab:table2}).
\begin{table}
\caption{Parameters involved in the fits of $^{28}$Si$^{32}$S lines}
\label{tab:table2}
\begin{tabular}{ccccc}
\hline
Parameter & Units & Value & Error & Ref.\\
\hline
$x(1\rstar\le r<\rdin)$          & $\times 10^{-6}$ & $4.9$                & $+0.8/-0.6$   & 1 \\
$x(r=\rdout)$                     & $\times 10^{-6}$ & $1.6$                & $+0.3/-0.3$   & 1 \\
$x(r\ge 50\rstar)$                & $\times 10^{-6}$ & $1.3$                & ---           & 2 \\
$T_\subscript{rot}(\rstar)$         & K               & $2330$               & ---           &  \\
$T_\subscript{rot}(\rdin)$          & K               & $860$                & $+130/-120$   & 1 \\
$T_\subscript{rot}(\rdout)$         & K               & $400$                & $+70/-40$    & 1 \\
$\gamma_\subscript{rot}$            &                 & $0.55$               & ---           & 2\\
$T_\subscript{vib,$v$=1-0}(\rstar)$ & K               & $1560$               & $+140/-120$   & 1 \\
$T_\subscript{vib,$v$=1-0}(\rdin)$ & K                & $613$                & $+11/-11$     & 1 \\
$T_\subscript{vib,$v$=1-0}(\rdout)$ & K               & $227$                & $+12/-12$     & 1 \\
$T_\subscript{vib,$v$=2-1}(\rstar)$ & K               & $1290$               & $+130/-80$    & 1 \\
$T_\subscript{vib,$v$=2-1}(\rdin)$  & K               & $490$                & $+18/-10$     & 1 \\
$T_\subscript{vib,$v$=2-1}(\rdout)$ & K               & $190$                & $+60/-70$     & 1 \\
$T_\subscript{vib,$v$=3-2}(\rstar)$ & K               & $760$                & $+80/-100$    & 1 \\
$T_\subscript{vib,$v$=3-2}(\rdin)$  & K               & $418$                & $+20/-20$     & 1 \\
$T_\subscript{vib,$v$=3-2}(\rdout)$  & K              & $153^*$         & ---           & 1 \\
$T_\subscript{vib,$v$=4-3}(\rstar)$ & K               & $550$                & $+220/-230$   & 1 \\
$T_\subscript{vib,$v$=4-3}(\rdin)$  & K               & $313^*$         & ---          & 1 \\
$T_\subscript{vib,$v$=4-3}(\rdout)$  & K              & $117^*$         & ---           & 1 \\
$T_\subscript{vib,$v$=5-4}(\rstar)$ & K               & $312^\dagger$               &              & 1 \\
$T_\subscript{vib,$v$=5-4}(\rdin)$  & K               & $224^*$         & ---           & 1 \\
$T_\subscript{vib,$v$=5-4}(\rdout)$  & K              & $80^*$          & ---           & 1 \\
$\gamma_\subscript{vib}$            &                 & $1.0$                & ---           & 3 \\
\hline
\end{tabular}
\newline
$x$: abundance relative to H$_2$;
$T_\subscript{rot}$: rotational temperature;
$\gamma_\subscript{rot}$: exponent of the rotational 
temperature power-law outwards from \rdout;
$T_\subscript{vib}$: vibrational temperature;
$\gamma_\subscript{vib}$: exponent of the vibrational 
temperature power-law outwards from \rdout{} 
for all the vibrational bands.
$T_\subscript{rot}$ and $T_\subscript{vib}$ are $\propto r^{-\gamma}$, where
$\gamma$ depends on the Region of the CSE and the temperature.
The parameters for which no uncertainty has been provided (---) were assumed
as fixed throughout the whole fitting process.
$^*$Linearly extrapolated from the vibrational temperature of other 
bands at the same position in the envelope.
$^\dagger$This parameter is free but the fitted lines are too weak to derive a 
reliable error.
References: 
(1) This work
(2) \citet{agundez_2012}
(3) \citet{fonfria_2008}.
\end{table}
The parameters related to $^{28}$Si$^{32}$S derived from the fits were the 
abundance with respect to H$_2$, the vibrational temperature at the stellar 
photosphere, and the rotational and vibrational temperatures at the 
inner and outer acceleration shells.
The $^{28}$Si/$^{29}$Si, $^{28}$Si/$^{30}$Si, and $^{32}$S/$^{34}$S isotopic ratios 
were estimated from the lines of $^{29}$Si$^{32}$S, $^{29}$Si$^{32}$S, and 
$^{28}$Si$^{34}$S as scaling factors of the abundance profile of $^{28}$Si$^{32}$S
(Section~\ref{sec:results.isotopologues}).
The fits to the lines of the SiS isotopologues were performed assuming the 
rotational and vibrational temperatures derived from the fits to the lines 
of $^{28}$Si$^{32}$S.
Several other parameters related to the dust component of the envelope, such as 
the dust optical depth or the temperature of the dust grains, were left as 
free parameters in order to fit the continuum and kept fixed when
the SiS lines were fitted.
The fits were carried out using all the observed lines from each isotopologue 
simultaneously.
A number of observed SiS lines were slightly offset along the 
frequency and normalised flux level axes to get the best fit.
These offsets in the normalised flux level 
axis are included to remove artefacts 
introduced during the baseline removal or due to blends with other 
lines coming from \irc{} or weak telluric lines.
The offsets in the frequency axis are always smaller than the spectral 
resolution of the observations ($\simeq 3-4$~\kms) but larger than typical 
with TEXES observations.
The offsets have larger amplitude, although still generally less than
2~\kms, toward higher frequencies.
The line list we used is based on experimental measurements and
has small errors in our spectral range ($\lesssim 0.5$~\kms), 
as expected for a simple molecule like SiS.
We see slightly small amplitude frequency offsets 
when we looked at the atmospheric features, suggesting that the determination
of the frequency scale is less precise at higher frequencies for these data.

\begin{figure}
\includegraphics[width=0.475\textwidth]{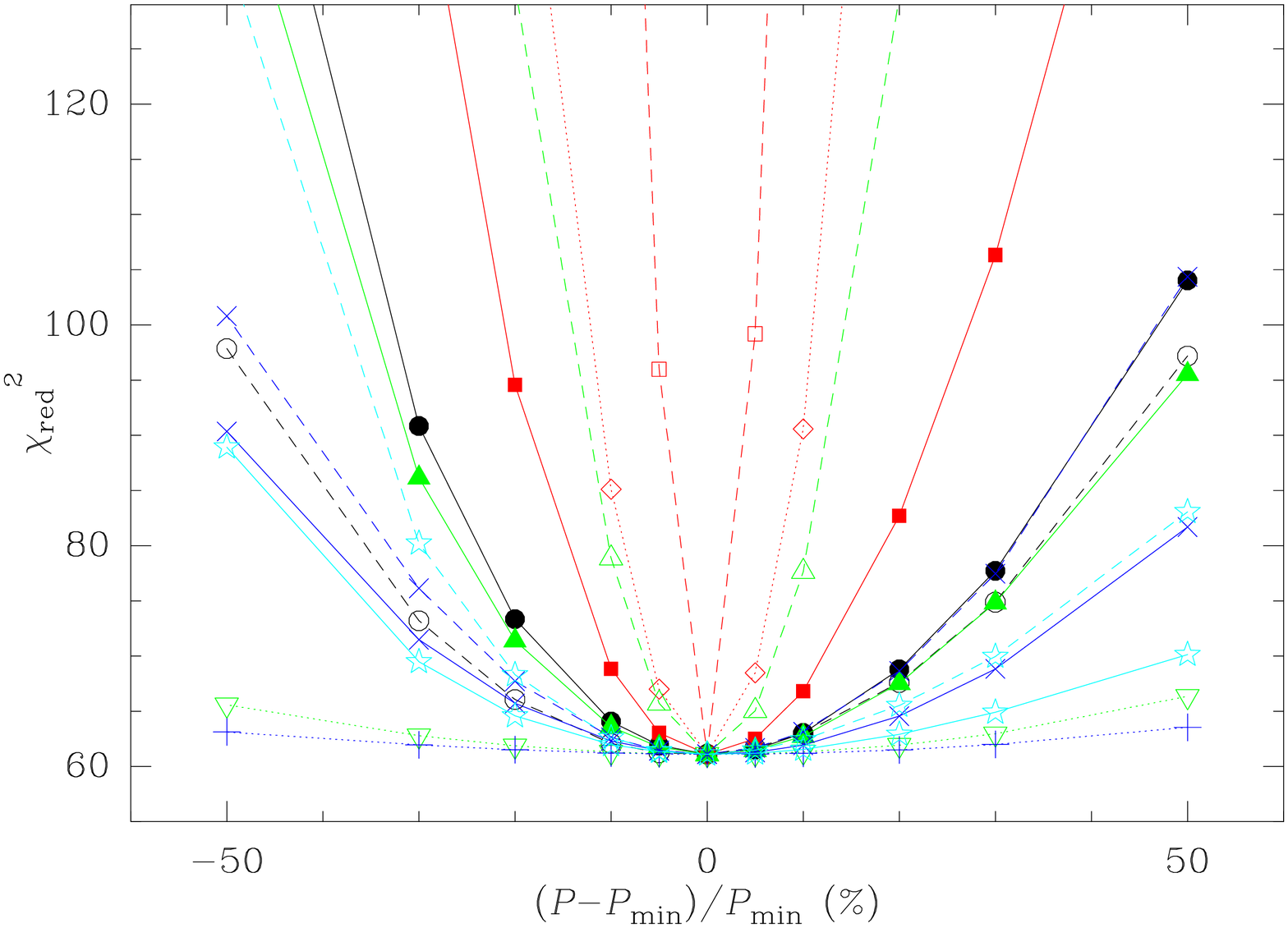}
\caption{Dependence of the $\chi_\subscript{red}^2$ function around its minimum  
($\chi_\subscript{red,min}^2\simeq 60.8$) on a given free parameter $P$.
$P_\subscript{min}$ is the value of the parameter $P$ derived from the best fit.
The parameter $x(\rstar\le r\le\rdin)$ is plotted in solid black ($\bullet$), 
$x(r=\rdout)$ in dashed black ($\circ$), 
$T_\subscript{rot}(\rdin)$ in solid cyan ($\star$), 
$T_\subscript{rot}(\rdout)$ in dashed cyan ($\star$), 
$T_\subscript{vib,$v=1-0$}(\rstar)$ in solid red ($\blacksquare$),
$T_\subscript{vib,$v=1-0$}(\rdin)$ in dashed red ($\square$), 
$T_\subscript{vib,$v=1-0$}(\rdout)$ in dotted red ($\diamond$),
$T_\subscript{vib,$v=2-1$}(\rstar)$ in solid green ($\blacktriangle$), 
$T_\subscript{vib,$v=2-1$}(\rdin)$ in dashed green ($\vartriangle$), 
$T_\subscript{vib,$v=2-1$}(\rdout)$ in dotted green ($\triangledown$), 
$T_\subscript{vib,$v=3-2$}(\rstar)$ in solid blue ($\times$),
$T_\subscript{vib,$v=3-2$}(\rdin)$ in dashed blue ($\times$),
and $T_\subscript{vib,$v=4-3$}(\rstar)$ in dotted blue ($+$).
See the caption of Tables~\ref{tab:table1} and \ref{tab:table2} for the 
definitions of these parameters.
}
\label{fig:f2}
\end{figure}

The model was sensitive to changes in most of the parameters related to SiS 
at distances $\lesssim 35$\rstar{} and to dust up to $\simeq 400$\rstar.
The uncertainties of the free parameters 
(Tables~\ref{tab:table1} and \ref{tab:table2}) 
were calculated by varying them until the 
the difference between the observed and the computed synthetic 
spectrum was larger than the detection limit of $3\sigma$ in each channel.
This conservative decision directly affects the derived uncertainties,
which could be overestimated.
During this process, all free parameters were varied at the same time
to account for any possible influence between them that could substantially 
affect the derived uncertainties.
These errors reflected the sensitivity of the model to each parameter,
as can be seen in the shape of the $\chi_\subscript{red}^2$ 
function (Fig.~\ref{fig:f2}), which shows the largest 
curvature around its minimum (best fit) for the most influential parameters 
on the model (e.g., the vibrational temperature of bands $v=1-0$ and $2-1$ 
at the end of the inner acceleration shell, \rdin).
The minimum $\chi_\subscript{red}^2$ achieved during the best fit is $\simeq 60$.
This large value is mostly related to contamination of the SiS lines 
by other features in the
frequency intervals used to calculate the $\chi_\subscript{red}^2$ function.
Although these differences do not severely affect the global
minimisation process due to the large number of lines involved, 
they increase the minimum value of $\chi_\subscript{red}^2$.
This value could be diminished by modifying the frequency interval
where the $\chi_\subscript{red}^2$ 
function is calculated for each line to remove what seems a 
contamination \textit{after} the end of the fitting
process, but this choice could compromise the objectivity of the 
minimisation procedure and was discarded.

\section{Results and analysis}
\label{sec:results}

\subsection{$^{28}$Si$^{32}$S}
\label{sec:results.sis}

Most of the observed lines in bands $v=1-0$ and $2-1$ show shapes compatible
with P-Cygni profiles (Fig.~\ref{fig:f1}).
The emission component cannot be clearly recognised in the profiles of the 
lines of band $v=3-2$, although this fact can be consequence of strong overlaps 
with close lines.
This component is insignificant in the profiles of the lines of bands 
$v=4-3$ and $5-4$.
The lines of band $v=1-0$ display typical absorption depths of $10-20$~per~cent
of the continuum emission while the lines of bands $v=2-1$, $3-2$, $4-3$, and 
$5-4$ are weaker with absorption depths smaller than 5, 2.5, 1.3, and 
0.7~per~cent of the continuum, respectively.
The maximum absorption for all the lines occurs at velocities between $-14.3$ 
and $-4.5$~\kms.
The absorption component of the lines of band $v=1-0$ with 
$J_\subscript{low}\lesssim 12$ and $\gtrsim 23$ show main contributions 
peaking at $\simeq -13.5$ and $-9$~\kms{} in average, respectively.
The absorption component of these lines have widths of $\simeq 7$~\kms,
even for the lines involving ro-vibrational levels with the highest energies.
The absorption of the lines with $12\lesssim J_\subscript{low}\lesssim 23$
show both absorption contributions (at $-13.5$ and $-9$~\kms) and,
consequently, the largest absorption widths ($\simeq 9-10$~\kms).
A width of 7~\kms{} is also found in the absorption components of the lines of
band $v=2-1$, which display their maximum between $-10$ and $-5$~\kms. 
The absorption of the lines of bands $v=3-2$, $4-3$, and $5-4$ show smaller 
widths of $\simeq 2.5-6.5$~\kms, peaking between $-6$ and $-4$~\kms.

The velocity of the maximum absorption and the shape of the absorption
component of the lines are consequences of the expansion velocity of the
emitting gas.
The width of the absorption components can be produced by expansion velocity 
gradients, line broadening due to turbulence, the varying thermal line width 
due to the large temperature gradient, and 
the different directions along what the absorbing gas expands (more important 
for the gas close to the star).
Thus, the absorption of the low excitation lines is mostly formed far from 
the star where the gas is colder and it expands at large velocities.
The high excitation lines require warm gas and/or an efficient radiative 
pumping mechanism, conditions that are fulfilled close to the star.
Hence, the absorption of the observed high excitation lines indicate that the 
gas in the innermost envelope expands at low velocities.
The observed medium excitation lines, mostly produced between the regions
where the low and high excitation lines are formed, display large absorption 
widths which suggest the existence of an expansion velocity gradient due to 
the absorption of gas at different velocities.
The detailed analysis of the absorption components of the observed lines 
performed in the current work suggests
that the actual expansion velocity profile and our choice (Section~\ref{sec:model})
are essentially compatible but it can be improved close to the star.

\subsubsection{Improving the gas expansion velocity field}
\label{sec:velocity}

\begin{figure}
\includegraphics[width=0.475\textwidth]{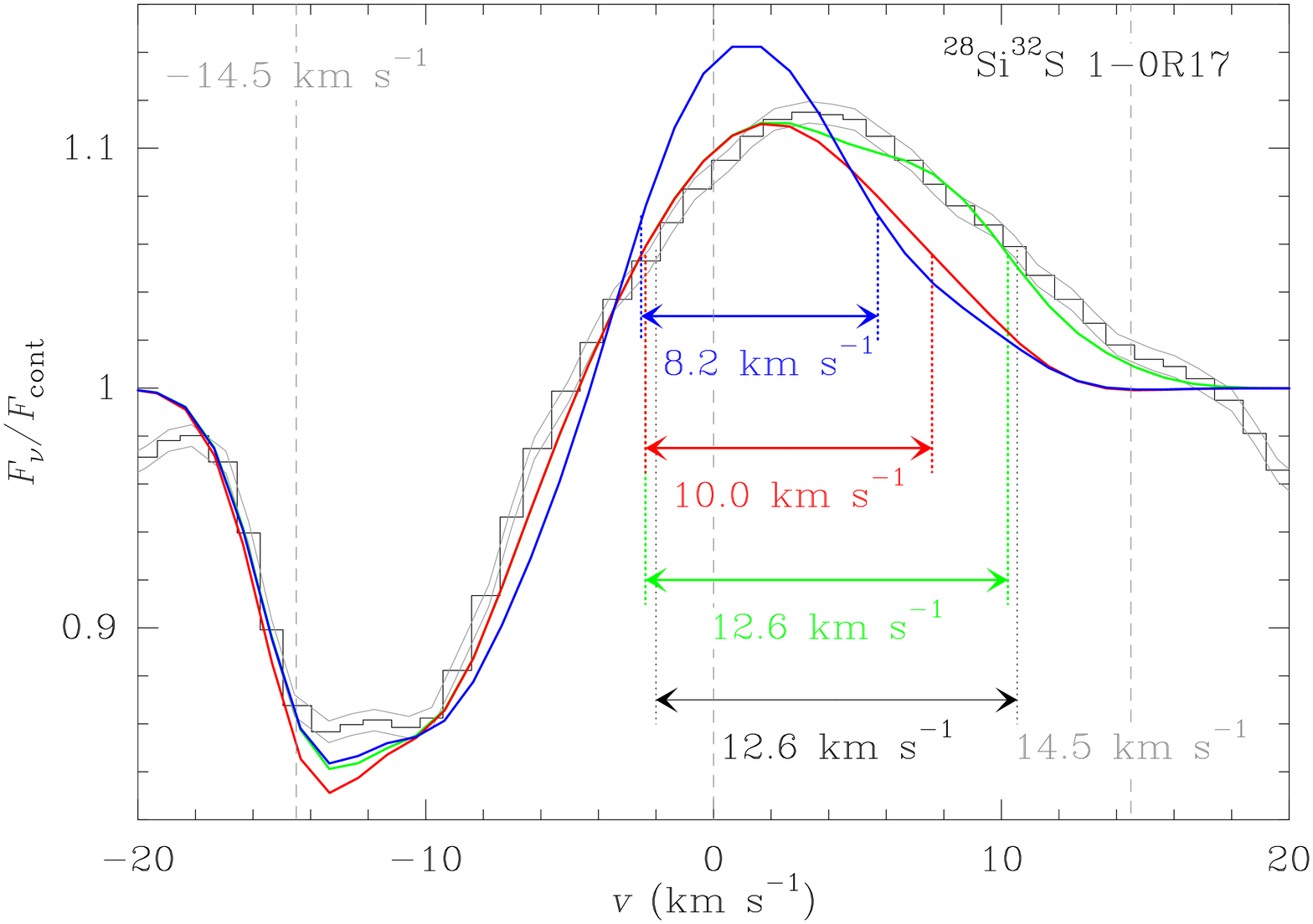}
\caption{Comparison between the observed $^{28}$Si$^{32}$S line 
$1-0\textnormal{R}(17)$ (black histogram) and the synthetic line calculated
with three different models 
depending on the expansion velocity field in Region I (blue and red),
and on the symmetry of the envelope (green).
All these lines have been taken from the best global fit to the observations.
The line in blue has been calculated with a constant expansion velocity 
of 5~\kms{} in Region I.
The line in red is produced by a spherical symmetric envelope with
a gas expansion velocity of $1+2.5(r/\rstar-1)$ between 1\rstar{}
and 5\rstar{} (Region I).
The line in green is the synthetic line calculated assuming the envelope
model of the red line but with an asymmetric vibrational
temperature distribution (see Section~\ref{sec:redshifted.emission}).
The curves in grey depart $3\sigma$ from the observations, where $\sigma$ 
is the rms of the noise of the spectrum ($\simeq 0.15$~per~cent of the 
continuum).
The widths included in the Figure stand for the FWHM of the emission 
component of the profiles.}
\label{fig:f3}
\end{figure}

The part of the synthetic SiS line profiles arising
from the gas in Regions II and III is in a reasonably good agreement with the 
observations. 
These calculations assumed the expansion velocity field initially considered 
in our model (Section~\ref{sec:model}).
Although a constant gas expansion velocity of 5~\kms{} 
can be adopted as the average velocity in Region I 
($1\rstar\le r<5\rstar$),
a linear dependence between the stellar atmosphere and 5\rstar{} ranging
from 1 to 11~\kms, i.e.,
$v_\subscript{exp}(1\rstar\le r<5\rstar)=1+2.5(r/\rstar-1)$~\kms,
describes better the emission component of 
the $^{28}$Si$^{32}$S lines of bands $v=1-0$ and $2-1$ (Fig.~\ref{fig:f3}).
This velocity gradient gives rise to a wider emission
component of the low excitation lines, although it is worth noting that
the observed lines of band $v=1-0$ with $J\lesssim 40$
show a red-shifted emission excess with respect to the synthetic lines that
is not accounted by the improved expansion velocity field
(see Figs.~\ref{fig:f1} and \ref{fig:f3}, 
and Section~\ref{sec:redshifted.emission} for
a discussion about this topic).
Lines of bands $v=3-2$, $4-3$, and $5-4$ are reasonably well reproduced with
both expansion velocity fields since these lines are formed close to the
photosphere \citep{agundez_2012,velilla_2015}, 
where the expansion velocity is expected to be low.

Hence, we used the improved expansion velocity field with the gradient
in Region I in our fits.

\subsubsection{$^{28}$Si$^{32}$S abundance with respect to H$_2$}
\label{sec:abundance}

\begin{table}
\caption{Abundance of $^{28}$Si$^{32}$S relative to H$_2$}
\label{tab:table3}
\begin{tabular}{ccccccc}
\hline
$\rstar$ & $\rdin$ & $\rdout$ & $\ge 50\rstar$ & Error & IR Phase & Ref.\\
\hline
4.9 & 4.9 & 1.6 & 1.3 & $^*$ & 0.08 & 1 \\
3.0 & 3.0 & 2.4 & 1.3 & $1.5-2.0$ & Multiepoch & 2 \\
4.0 & 4.0 & 4.0 & 4.0 & $4$ & 0.04 & 3 \\
19  & 17  & 1.7 & 1.7 & $2-3$ & $0.05-0.24$ & 4 \\
51 & 33 & 2.5 & $\le 0.9$ & $2$ & 0.5 & 5 \\
7.5 & 7.5 & 7.5 & 7.5 & $2-3$ & $0.09-0.29$ & 6 \\
\hline
\end{tabular}
\newline
The abundances are multiplied by $10^{-6}$.
Their errors are provided as multiplicative/dividing factors.
The pulsation phase was calculated following the analysis of the
infrared light curves performed by \citet{monnier_1998}.
$^*$See Table~\ref{tab:table2}.
(1) This work
(2) \citet{agundez_2012}
(3) \citet{decin_2010}
(4) \citet{schoier_2007}
(5) \citet{boyle_1994}
(6) \citet{bieging_1989}
\end{table}

The best fit to the whole set of lines derived assuming the envelope model 
presented above was achieved with a constant $^{28}$Si$^{32}$S abundance with 
respect to H$_2$ by number of $4.9\times 10^{-6}$ in Region I followed by a 
linear decrease in Region II down to $1.6\times 10^{-6}$ at the outer 
acceleration shell, and a further decrease outwards down to $1.3\times 10^{-6}$ 
at 50\rstar.
The $^{28}$Si$^{32}$S abundance beyond the outer acceleration shell is 
compatible with the results by \citet{boyle_1994}, \citet{schoier_2007}, and
\citet{agundez_2012} within a factor of 2 and with the results by 
\citet{decin_2010} within up to a factor of 3, a disagreement smaller than the 
uncertainty they estimated for their own results (Table~\ref{tab:table3}).
However, our $^{28}$Si$^{32}$S abundance is about 10 and 4 times smaller in 
Region I than the results by \citet{boyle_1994} and \citet{schoier_2007}, 
respectively.
The large discrepancy of one order of magnitude found between our derived 
abundance in Region I and that proposed by \citet{boyle_1994} is due to 
significant differences in the spectra.
In the spectrum presented by \citet{boyle_1994} only the blue-shifted part of 
the lines, corresponding to the absorption depths, were totally covered during 
the observations while the emission components were largely (but not completely) 
missed.
The emission component of most of the lines they observed was probably
blended with telluric lines or features of other molecules existing in the
envelope of \irc.
In addition, it is usually difficult to determine the baseline 
in spectral regions with many molecular lines, such as the region containing
SiS \citep*[see figs.~1-6 in][]{fonfria_2008}.
Thus, the line profiles were assumed to display no emission resulting in lines 
with absorption components deeper than they actually were due to the remnants 
of the emission components.
The modelling of these lines required an exceedingly low vibrational 
temperature and a
significant increase of the estimated $^{28}$Si$^{32}$S abundance in Region I 
and part of Region II compared to the actual profile. 
\citet{schoier_2007} derived the abundance by fitting previous observations 
acquired with BIMA by \citet{bieging_1993}.
These observations seem not to be affected by any instrumental effect
suggesting that the disagreement between our results and those by
\citet{schoier_2007} may be produced by an actual variation in the 
$^{28}$Si$^{32}$S emission.
Such variation would not be related to the stellar pulsation phase,
known to influence the emission of several molecules
\citep{carlstrom_1990,cernicharo_2014}, 
since both observations were carried out close to maximum light
\citep*[see Table~\ref{tab:table3};][]{monnier_1998}.
The abundances proposed by \citet{bieging_1989} and 
\citet{decin_2010}, although based on observations acquired near
maximum light, are affected by large errors that prevent us from deriving 
reliable conclusions about the time evolution.
Therefore, the observed variation in the $^{28}$Si$^{32}$S abundance
could be the consequence of a larger time-scale change in the mass-loss rate
\citep{cernicharo_2015a,cernicharo_2015c} or the gas phase chemistry close to 
the star, considering that our observations were acquired about 14~yr after the
observations taken by \citet{bieging_1993}, and the ejected gas is expected to 
spend about $10-15$~yr travelling across Region I.

The $^{28}$Si$^{32}$S abundance with respect to H$_2$ calculated with chemical 
models under thermodynamical equilibrium (TE) is larger than $\simeq 10^{-5}$ 
in Region I \citep{willacy_1998,agundez_2006}, i.e., a factor 
of 2 larger than our results.
The TE and chemical kinetic models give constant abundances in most of Region I 
and beyond, meaning that gas-phase chemical reactions involving SiS are only 
expected to play an important role in the vicinity of the star
\citep*[e.g.,][]{agundez_2006}.
Hence, the large decrease in the SiS abundance outwards from the inner
acceleration zone is probably due to depletion on to dust grains, as was 
suggested previously 
\citep{turner_1987,bieging_1989,bieging_1993,boyle_1994,schoier_2007,
agundez_2012}.

\subsubsection{Rotational and vibrational temperatures}
\label{sec:temperatures}

\begin{figure}
\centering
\includegraphics[width=0.475\textwidth]{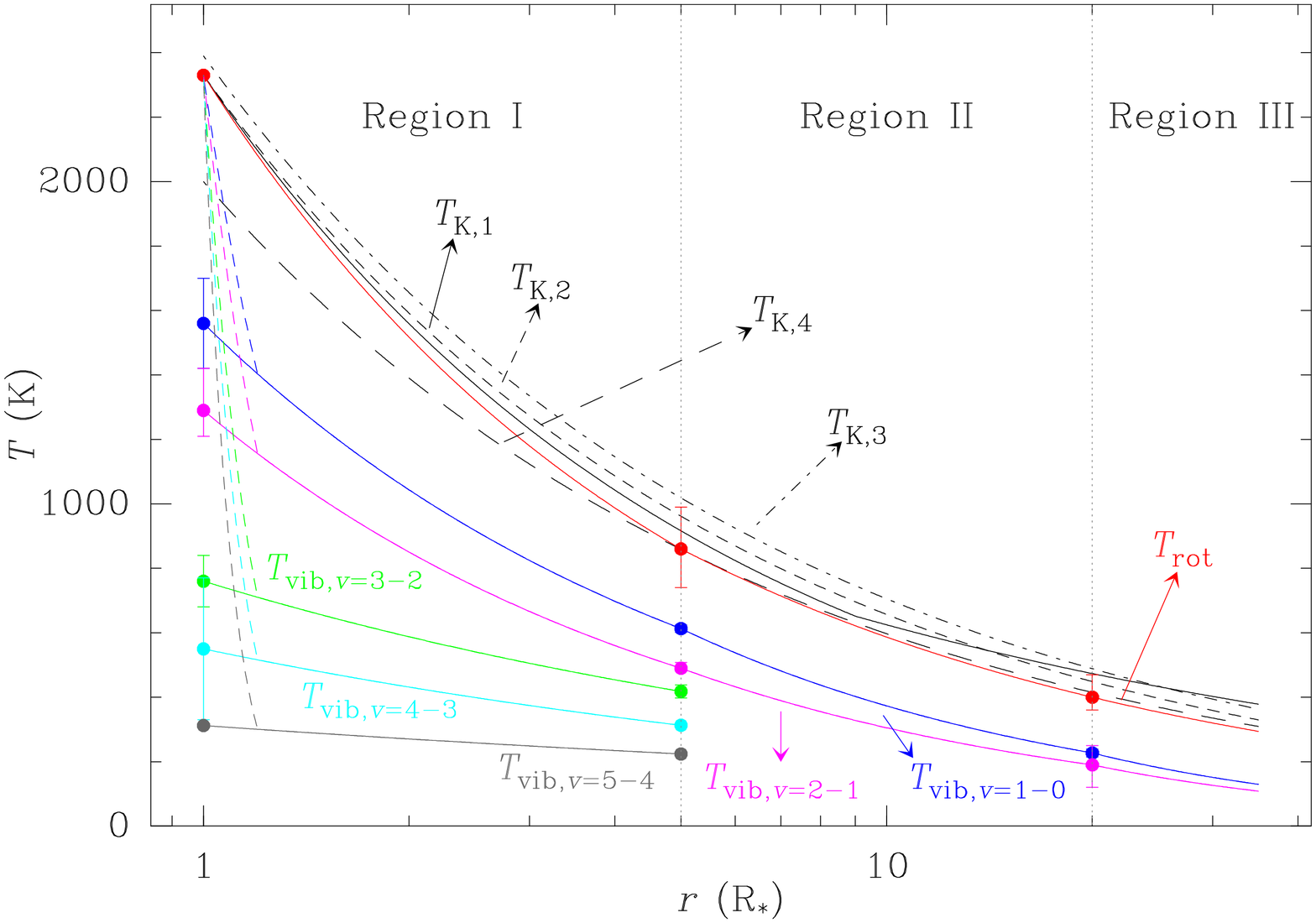}
\caption{Kinetic, rotational, and vibrational temperatures of $^{28}$Si$^{32}$S
from the stellar photosphere up to 35\rstar.
The kinetic temperatures, $T_\subscript{k}$, derived in previous works 
(references 1, 2, 3, and 4) are plotted in solid, short dashed, dashed-dot, 
and long dashed, black lines.
The rotational temperature, $T_\subscript{rot}$, derived in the frame of the 
current work is plotted in solid red.
The vibrational temperatures, $T_\subscript{vib}$, derived from ro-vibrational 
bands $v=1-0$, $2-1$, $3-2$, $4-3$, and $5-4$ are plotted in blue, magenta, 
green, cyan, and grey solid lines, respectively.
The dashed lines in these colours represent possible dependences of the
vibrational temperatures on the distance to the star between the stellar 
photosphere and $\simeq 1.2$\rstar, where \citet{cernicharo_2013} propose a 
larger gas density than we assume in the current work.
$T_\subscript{vib,$v=3-2$}$, $T_\subscript{vib,$v=4-3$}$, and 
$T_\subscript{vib,$v=5-4$}$ are only plotted up to 5\rstar{} because the lines
of these bands are mostly formed in Region I and no information on larger 
distances to the star can be retrieved from them.
References:
(1) \citet{debeck_2012}
(2) \citet{agundez_2012}
(3) \citet{schoier_2007}
(4) \citet{boyle_1994}
}
\label{fig:f4}
\end{figure}

The rotational temperature profile in the ground vibrational state derived from 
our fits throughout the dust formation zone 
\citep*[$1\rstar\le r\lesssim 30\rstar$; e.g.,][]{fonfria_2008,decin_2010}
are similar to the kinetic 
temperature proposed in previous works within the estimated uncertainties 
(Fig.~\ref{fig:f4}), demonstrating that the $^{28}$Si$^{32}$S molecules in the 
vibrational ground state is under rotational LTE throughout the first 
$\simeq 35$\rstar{} of the envelope.
Moreover, our results suggest that the rotational temperature for the 
vibrational states $v=1$ to 5 can be assumed to be the same for the vibrational 
ground state in the region of the envelope where these vibrational states are 
significantly populated, i.e., at distances to the star $\lesssim 10$\rstar.

The vibrational temperatures derived from bands $v=1-0$, $2-1$, $3-2$, $4-3$, 
and $5-4$ are significantly smaller than the kinetic temperature 
(Fig.~\ref{fig:f4}), which indicates that $^{28}$Si$^{32}$S is vibrationally 
out of LTE and its vibrational states are, mainly, radiatively populated.
Using the gas density profile presented in Section~\ref{sec:model}, we 
estimate the gas density at the stellar photosphere, 
$n_{\subscript{gas}}(\rstar)$, to be
$\simeq 4.2\times 10^{10}$~cm$^{-3}$.
The critical density, $n_\subscript{crit}$, at this position can be estimated 
with the A-Einstein coefficients of the observed $^{28}$Si$^{32}$S
ro-vibrational transitions, which range from 0.8 to 6.2~s$^{-1}$
\citep*[the MADEX code;][]{cernicharo_2012,velilla_2015},
and the de-excitation ro-vibrational state-to-state collisional rates 
discussed in \citet{velilla_2015}, where the calculations
performed by \citet{tobola_2008}
for the colliding pair $^{28}$Si$^{32}$S:He
were improved and extrapolated to high ro-vibrational levels.
As far as we know, there is no data in the literature about ro-vibrational 
transitions due to collisional processes between $^{28}$Si$^{32}$S and H$_2$.
Thus, we assume that the collisional rates are the same for H$_2$ and for He,
and that these rates, calculated for a kinetic temperature ranging between 
300 and 1500~K, can be extrapolated up to $\simeq 2300$~K following 
the power-law $T_\subscript{k}^{1/2}$.
We estimate that the maximum $^{28}$Si$^{32}$S ro-vibrational collisional rates 
ranged from $1.8\times 10^{-13}$ to $1.5\times 10^{-12}$~cm$^3$~s$^{-1}$ for the 
transitions of bands $v=1-0$ to $5-4$.
Hence, $n_\subscript{crit}\gtrsim 4\times 10^{12}$~cm$^{-3}$ at the stellar 
photosphere, two orders of magnitude larger than the estimated gas density.
However, the analysis of the HNC emission performed by \citet{cernicharo_2013} 
is compatible with an H$_2$ density very close to the star larger than 
$10^{12}-10^{13}$~cm$^{-3}$ as a consequence of the matter levitation due 
to the stellar pulsation 
\citep{bowen_1988,willacy_1998,agundez_2006,agundez_2012,cherchneff_2006},
being similar or exceeding the estimated critical density
($n_\subscript{gas}(\rstar)\gtrsim n_\subscript{crit}$).
Thus, the vibrational excitation would be dominated by collisions and
$^{28}$Si$^{32}$S might be vibrationally under or close to LTE in the vicinity 
of the stellar photosphere, meaning that the vibrational temperatures at this
position we derive in the current work are significantly constrained by the 
power law consider to prevail throughout Region I and should be assumed as 
lower limits.
In this case, a more realistic approach would include a new Region adjoining 
the photosphere of less than 1\rstar{} thick (e.g., an extension of the 
photosphere), where the dependence of the vibrational temperature with
the distance to the star would be much steeper than in the rest of Region I 
(Fig.~\ref{fig:f4}).
The larger flatness in Region I showed by $T_\subscript{vib,$v=3-2$}$,
$T_\subscript{vib,$v=4-3$}$, and $T_\subscript{vib,$v=5-4$}$ in comparison to the 
vibrational temperature derived from bands $v=1-0$ and $2-1$ are probably 
related to the lack of reliable information about the emission component of 
the lines of bands $v=3-2$, $4-3$, and $5-4$.
Thus, the vibrational temperature for these bands could be steeper than our 
best fit suggests throughout Region I,
as the results by \citet{velilla_2015} indicate.
On the other hand, the decrease experienced by the vibrational temperature 
in Region I when the vibrational quantum number $v$ increases 
($T_\subscript{vib,$v=1-0$}(\rstar)$ compared to 
$T_\subscript{vib,$v=2-1$}(\rstar)$, $T_\subscript{vib,$v=3-2$}(\rstar)$, and so 
on) derived from our fits seems to be significant, as the uncertainties suggest 
(Table~\ref{tab:table2}).
This variation is probably caused by the different A-Einstein coefficients 
of the ro-vibrational transitions of bands $v=1-0$ to $5-4$, that are about 
$v_\subscript{up}$ times larger than for band $1-0$.

\subsubsection{The red-shifted emission excess}
\label{sec:redshifted.emission}

The observed lines of the $v=1-0$ band with $J_\subscript{low}\lesssim 40$ show a 
red-shifted emission excess at Doppler velocities of $\gtrsim 6-10$~\kms{}
with respect to the lines derived from the best
fit.
The affected lines involve ro-vibrational levels with energies below 700~K,
suggesting that the emitting gas is located 
further than $\simeq 15\rstar$ from the star.
This scenario is compatible with the lack of an emission excess in higher
excitation lines, which are mostly produced closer to the star than 
$\simeq 10\rstar$.
A similar 
red-shifted emission excess was already reported by \citet{fonfria_2008}
in the ro-vibrational lines of C$_2$H$_2$ and HCN.
Interferometer observations carried out with HPBW~$\simeq 0.25-0.35$~arcsec
of the lines Si$^{34}$S($v=0,J=15-14$), H$^{13}$CN($v=0,J=3-2$), and 
SiO($v=0,J=6-5$) \citep{shinnaga_2009,fonfria_2014} also showed 
red-shifted emission excesses probably produced in the dust formation 
zone \citep*[$1\rstar\le r\lesssim 30\rstar$; e.g.,][]{fonfria_2008,decin_2010},
the same region where the 
ro-vibrational lines presented and analysed in the current work and
by \citet{fonfria_2008} are formed.

A spherically symmetric envelope model cannot reproduce the 
red-shifted emission excess
and the absorption component of many lines at the same time.
Since these absorption components highly constrain the possible structure,
the expansion velocity field, and the physical and chemical
conditions prevailing throughout the envelope, other models should be invoked
to explain the observations.
As most of the emission component of the SiS lines of the $v=1-0$ band comes 
from the rear part of the envelope close to 
the star while the absorption component is formed in front of 
the star, the red-shifted emission excess 
could be modelled assuming an asymmetric SiS abundance 
and/or vibrational temperature distributions.
This asymmetry could be a consequence of (1)
gas expanding at velocities higher than expected in the surroundings of 
the stellar photosphere through, e.g., the existence of an outflow behind the 
star \citep{fonfria_2014}, (2)
the influence of the companion on the ejected matter by the AGB star
by its orbital movement \citep{cernicharo_2015a,cernicharo_2015c},
or (3) an anisotropic mass loss from the AGB star.

A simple 2D envelope model can be proposed and its molecular
emission calculated with our code \citep{fonfria_2014}
in order to show the feasibility of an asymmetric emission.
In this model, we defined a conical region behind the star
with its axis parallel to the line-of-sight, its vertex matching up with
the centre of the star, 
and an aperture of 120$^\circ$, where the SiS abundance and the
vibrational temperature for band $v=1-0$ could be different than in the rest of
the envelope.
These magnitudes could only depend on the
distance to the star in both regions.
The aperture was chosen to modify the part of the emission component 
of the lines of the $v=1-0$ band where the excess was found.
The initial physical and chemical conditions were those derived from the
best fit assuming the spherically symmetric envelope 
(Tables~\ref{tab:table1} and \ref{tab:table2}).
In the best fit adopting the asymmetric model (see Fig.~\ref{fig:f3}),
the vibrational temperature of the $v=1-0$ band
in the conical region behind the star
was incremented from $\simeq 230$ up to $\simeq 300$~K 
at the outer acceleration shell (located at 20\rstar),
following the law $\propto r^{-0.45}$ outwards.
The variation of the SiS abundance in the conical region derived from
the fitting process was negligible.
This 2D envelope model did not introduce any significant modification in the
synthetic lines of bands $v=2-1$, $3-2$, $4-3$, and $5-4$ compared
to the results of the spherically symmetric envelope model.

\subsection{Isotopic ratios}
\label{sec:results.isotopologues}

\begin{table}
\caption{Isotopic ratios in the innermost envelope}
\label{tab:table4}
\begin{tabular}{c@{\hspace{8ex}}c@{\hspace{8ex}}c@{\hspace{8ex}}c}
\hline
Ratio & Value & Error & Reference\\
\hline
$^{28}$Si/$^{29}$Si  & 17     & $+5/-4$      & 1 \\
                   &  18     & $+2/-2$       & 2 \\
                   & 8.2     & $+1.7/-1.7$   & 3 \\
                   & 15.1    & $+0.7/-0.7$   & 4\\
                   & 17.2    & $+1.1/-1.1$   & 5\\
                   & $>15.4$ & ---           & 6 \\
$^{32}$S/$^{34}$S    & 14     & $+6/-4$        & 1 \\
                   &  $22.0$ & $+2.5/-2.5$   & 2 \\
                   & 30      & $+8/-8$       & 3 \\
                   & 19.6    & $+1.3/-1.3$   & 4\\
                   & 18.9    & $+1.3/-1.3$   & 5 \\    
                   & 21.8    & $+2.6/-2.6$   &6 \\
\hline
\end{tabular}
\newline
(1) This work
(2) \citet{agundez_2012}
(3) \citet{patel_2011}
(4) \citet{patel_2009}
(5) \citet{he_2008}
(6) \citet{kahane_2000} and \citet{cernicharo_2000}.
The isotopic ratios by \citet{patel_2011} presented here were calculated by 
assuming as reliable the isotopic ratios derived from optically thin lines
that do not show extended emission when possible (see table~6 of this 
reference).
\end{table}

The observed lines of $^{29}$Si$^{32}$S, $^{28}$Si$^{34}$S, and $^{30}$Si$^{32}$S 
belong to band $v=1-0$ and show shapes compatible with P-Cygni profiles 
(Fig.~\ref{fig:f1}).
These lines are weak, with differences between the maximum emission and 
absorption of about three to four times $\sigma$ for lines with 
$J_\subscript{low}\lesssim 50$.
Most of the lines in the observed frequency range are blended with C$_2$H$_2$, 
HCN, and $^{28}$Si$^{32}$S lines.
Only a handful of lines over the detection limit are sufficiently isolated
and can be conveniently fitted (Fig.~\ref{fig:f1}).
The lines of $^{29}$Si$^{32}$S and $^{28}$Si$^{34}$S
were strong enough to derive the isotopic ratios
$^{28}$Si/$^{29}$Si and $^{32}$S/$^{34}$S with a relatively good accuracy
of $\simeq 30$ and 40~per~cent, respectively (Table~\ref{tab:table4}).
The fit to the only $^{30}$Si$^{32}$S line suitable to be
analysed gave an isotopic ratio $^{28}$Si/$^{30}$Si 
of 39 with an error of 60~per~cent.
The lack of available lines and the large uncertainty
prevent us to derive reliable conclusions about this isotopic ratio.

The ratios $^{28}$Si/$^{29}$Si and $^{32}$S/$^{34}$S 
are compatible with previous estimates derived from observations
carried out in the mm wavelength range (Table~\ref{tab:table4}).
These observations mostly recovered the long scale emission of the envelope
that comes from layers composed of matter expelled more than $\simeq 800$~yr ago
\citep{cernicharo_2015c}, while our observations traced gas recently 
ejected.
Hence, no significant changes in the nucleosynthesis of the most
abundant Si and S isotopes
are expected to have happened during the last 1000~yr.

\section{Conclusions}
\label{sec:conclusions}

In this paper, we have presented high spectral resolution observations in 
the mid-IR range ($\simeq 12.5-14.0~\mu$m) towards the AGB star \irc.
Two hundred and four lines coming from band $v=1-0$, $2-1$, $3-2$, $4-3$, and
$5-4$ of $^{28}$Si$^{32}$S, 26 from $v=1-0$ of $^{29}$Si$^{32}$S, 20 from $v=1-0$ 
of $^{28}$Si$^{34}$S, and 15 from $v=1-0$ of $^{30}$Si$^{32}$S have been 
identified in the observed spectrum.
About 30~per~cent of the total set of detected lines are unblended or
partially blended and could be fitted and analysed.
We have found that:
\begin{itemize}
\item the quality of the fits is higher by assuming that the gas
expansion velocity profile grows linearly from the photosphere up to 5\rstar{}
following the dependence 
$v_\subscript{exp}(1\rstar\le r< 5\rstar)=1+2.5(r/\rstar-1)$~\kms{} compared 
to the fits based on a constant velocity of 5~\kms.
Our fits support the expansion velocity profile proposed by \citet{fonfria_2008}
outwards from 5\rstar, i.e., $v_\subscript{exp}(5\rstar\le r<20\rstar)=11$~\kms{} and
$v_\subscript{exp}(r\ge 20\rstar)=14.5$~\kms.
\item The abundance of $^{28}$Si$^{32}$S is $4.9\times 10^{-6}$ between the 
stellar photosphere and the inner acceleration shell (located at 5\rstar) and 
decreases linearly down to $1.6\times 10^{-6}$ at the outer acceleration shell
(at 20\rstar).
The observed lines are compatible with an abundance of $1.3\times 10^{-6}$ at 
50\rstar.
The decrease in the abundance beyond the inner acceleration shell is probably 
explained by the depletion of SiS on to the dust grains due to its refractory 
nature and the lack of gas-phase chemical reactions expected to be enabled
in this region of the CSE.
\item $^{28}$Si$^{32}$S is vibrationally out of LTE in most of the envelope.
The vibrational temperature for the vibrational ground state is
$1560(\rstar/r)^{0.58}$~K between the stellar surface and the inner 
acceleration shell (Region I) and $613(\rstar/r)^{0.72}$~K up to the outer 
acceleration shell (Region II).
For the $v=1$ vibrational state, the relation is $1290(\rstar/r)^{0.60}$~K 
in Region I and $490(\rstar/r)^{0.68}$~K in Region II.
Beyond the outer acceleration shell, the vibrational temperature for these
states is compatible with an exponent of $-1$.
For $v=2$, 3, and 4, the vibrational temperature is compatible with
$760(\rstar/r)^{0.37}$, $550(\rstar/r)^{0.35}$, and $312(\rstar/r)^{0.20}$~K 
in Region I, respectively.
\item $^{28}$Si$^{32}$S can be assumed to be rotationally under LTE with a 
rotational temperature of $2330(\rstar/r)^{0.62}$~K between the stellar 
photosphere and the inner acceleration shell and $860(\rstar/r)^{0.55}$~K 
outwards.
\item There is a red-shifted emission excess at large velocities in the 
$^{28}$Si$^{32}$S lines of bands $v=1-0$ and $2-1$ not noticed in the lines of 
bands $v=3-2$, $4-3$, and $5-4$.
This excess can be explained by an asymmetry of the vibrational temperature 
distribution for what the gas behind the star shows a higher excitation
around $\simeq 20\rstar$ that mostly affects the lines of band $v=1-0$.
\item The isotopic ratios $^{28}$Si/$^{29}$Si and 
$^{32}$S/$^{34}$S derived from our observations are 
17 and 14, respectively.
These ratios, referred to matter in the vicinity of the star, 
are compatible with the previously proposed values for outer shells of the
envelope.
\end{itemize}

\section*{Acknowledgements}

JPF was supported during part of this study by the UNAM through a postdoctoral 
fellowship.
We thank the Spanish MINECO/MICINN for funding support through grants
AYA2009-07304, AYA2012-32032, the ASTROMOL Consolider project CSD2009-00038
and the European Research Council (ERC Grant 610256: NANOCOSMOS).
MJR is supported by grant AST 03-07497. 
Development of TEXES was supported by grants from the NSF and USRA.
MJR, JHL, and others want to thank IRTF, which is operated by the University of 
Hawaii under Cooperative Agreement NCC 5-538 with the National Aeronautics and 
Space Administration, Office of Space Science, Planetary Astronomy Program.
MFL thanks the University of Illinois at Urbana-Champaign and the hospitality 
of the Instituto de Astronom\'ia (UNAM).
We thank the unknown referee for his/her helpful comments.

\appendix

\section{Gas expansion velocity field}
\label{sec:expansion.velocity}

In order to fit the observed lines, we \textnormal{initially}
chose the expansion velocity profile 
derived by \citet{fonfria_2008} from the analysis of several hundreds of 
C$_2$H$_2$ and HCN lines in the mid-IR.
During the current work we slightly modified this velocity profile
to improve the quality of the fits to the SiS lines proving that
the velocity profile derived by \citet{fonfria_2008} is essentially
compatible with the actual profile.
However, \citet{decin_2014} has recently published a new profile based on
the analysis of high angular resolution observations (HPBW~$\simeq 0.33$~arcsec) 
carried out with ALMA in the submillimetre wavelength range.
In this new profile, the gas expands at $2-3$~\kms{} between the stellar
surface and 5.6\rstar{} and grows linearly up to 11\rstar, where it reaches
the terminal expansion velocity (profile 1; not shown).
These authors also proposed a more complex profile compatible with their 
observations where the expansion velocity is 8~\kms{} between 8 and 10\rstar{} 
(profile 2; not shown).

\begin{figure}
\includegraphics[width=0.475\textwidth]{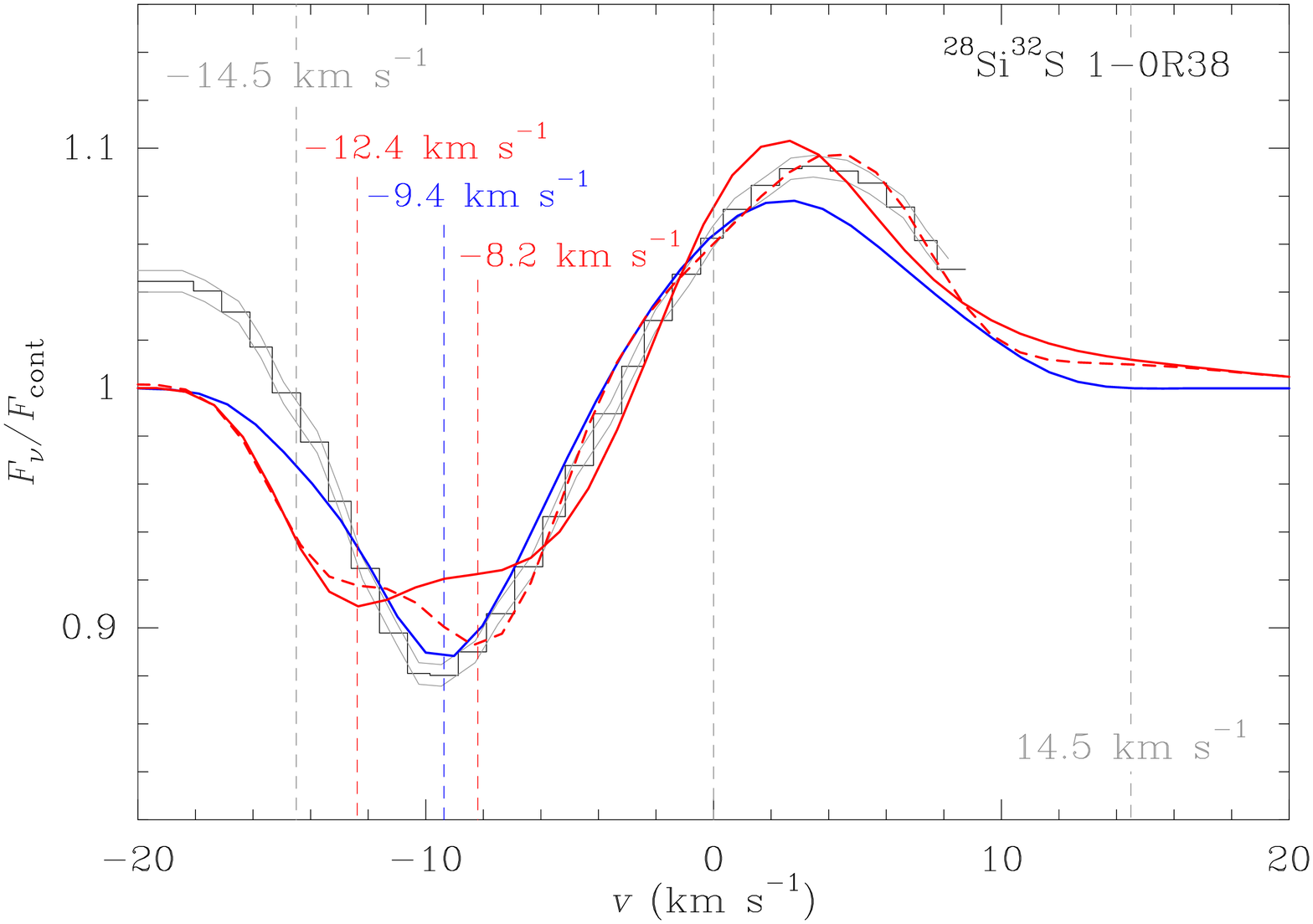}
\caption{Comparison between the observed $^{28}$Si$^{32}$S line 
$1-0\textnormal{R}(38)$ (black histogram) and the synthetic lines calculated 
with the expansion velocity profiles of \citet{decin_2014} (profile 1 and 
profile 2 in the text in solid and dashed red lines, respectively) and 
the current work (blue solid line).
The synthetic lines are derived from the best fits to the observed lines of 
band $v=1-0$.
The curves in grey depart $3\sigma$ from the observations, where $\sigma$ 
is the rms of the noise of the spectrum ($\simeq 0.15$~per~cent of the 
continuum).}
\label{fig:f5}
\end{figure}

We tried to fit our observations with these new velocity profiles finding
that it is not possible to get better fits than that the obtained with the 
profile derived in the current work (Section~\ref{sec:velocity}).
The largest deviations of the synthetic from the observed lines appear in the
lines of the ro-vibrational band $v=1-0$ with $J_\subscript{low}$ ranging from
30 to 50 (Fig.~\ref{fig:f5}).
The maximum absorption of these lines occurs between $-10$ and $-9$~\kms.
Assuming a spherically symmetric expanding envelope, the maximum absorption
is produced in front of the star by the shells of the envelope located at 
distances to it smaller than $\simeq 15-20$\rstar{} and where the overlapping 
fraction between the molecular absorption of contiguous shells is maximum.
A gradient in the expansion velocity field would diminish this fraction and,
consequently, the optical depth, making the absorption less prominent.
This diminishing could be balanced with a larger molecular abundance but
it would also affect the shape of the whole absorption component.
Adopting a continuous abundance profile described by its linearly connected 
values at 1, 5.6, and 11\rstar{} for the profile 1 and at 1, 5.6, 8, 10, and 
11\rstar{} for the profile 2, the best fits cannot properly reproduce the 
absorption component of the observed lines (Fig.~\ref{fig:f5}).
Both profiles produce absorptions for the lines of band $v=1-0$ with 
$J_\subscript{low}=30-50$ comprising two velocity contributions approximately at 
$-13$ and between $-9$ and $-7$~\kms, which are produced by the gas expanding 
at the terminal velocity and at $7-9$~\kms, respectively.
However, the minima of the absorption components of the observed lines are 
between both contributions.
The main characteristics of the absorption component of the observed lines are 
better explained with the expansion velocity field derived 
in the current work.

\end{document}